\documentclass[10pt, pre, twocolumn, showpacs, eps]{revtex4-1}
\usepackage{amsmath,amssymb,graphicx,color}
\newcommand{\dfracp}[2]{\dfrac{\partial #1}{\partial #2}}
\newcommand{\ave}[1]{\left\langle #1 \right\rangle}
\newcommand{\poisson}[2]{\left\{ #1 , #2 \right\}}

\begin{document}

\title{Non-diagonalizable and non-divergent susceptibility tensor
  in the Hamiltonian mean-field model
  with asymmetric momentum distributions}
\author{Yoshiyuki Y. Yamaguchi}
\email[E-mail: ]{yyama@amp.i.kyoto-u.ac.jp}
\affiliation{Department of Applied Mathematics and Physics, 
        Graduate School of Informatics, Kyoto University, 
        606-8501 Kyoto, Japan}
% 05.20.Dd Kinetic theory
% 05.70.Jk Critical point phenomena
% 74.25.N- Response to electromagnetic fields
%\today
\begin{abstract}
  We investigate response to an external magnetic field
  in the Hamiltonian mean-field model,
  which is a paradigmatic toy model of a ferromagnetic body
  and consists of plane rotators like the XY spins.
  Due to long-range interactions,
  the external field drives the system to a long-lasting
  quasistationary state before reaching thermal equilibrium,
  and the susceptibility tensor obtained in the quasistationary state
  is predicted by a linear response theory based on the Vlasov equation.
  For spatially homogeneous stable states,
  whose momentum distributions are asymmetric
  with zero-means,
  the theory reveals that the susceptibility tensor
  for an asymptotically constant external field 
  is neither symmetric nor diagonalizable,
  and the predicted states are not stationary accordingly.
  Moreover, the tensor has no divergence even at the stability threshold.
  These theoretical findings are confirmed by direct numerical
  simulations of the Vlasov equation
  for the skew-normal distribution functions.
\end{abstract}
\maketitle

\section{Introduction}
\label{sec:introduction}

Long-range Hamiltonian systems have many remarkable features
\cite{campa-dauxois-ruffo-09},
and one of them is existence of quasistationary states (QSSs)
in the way of relaxation to thermal equilibrium.
The lifetime of QSSs diverges with the number of particles
consisting of the system \cite{zanette-montemurro-03,yamaguchi-04},
and hence QSSs are solely observable in a system with large population
like self-gravitating systems \cite{binney-tremaine-08}.
Dynamics of such a system is described by the Vlasov equation,
or the collisionless Boltzmann equation,
in the limit of large population
\cite{braun-hepp-77,dobrushin-79,spohn-91},
and the QSSs, including thermal equilibrium states,
are regarded as stable stationary solutions to the Vlasov equation.
The system slowly goes towards thermal equilibrium
with large but finite population due to finite size effects
\cite{yamaguchi-04,barre-06}.

The QSSs are observed not only in isolated systems,
but also in systems under external fields.
The initial QSS, which may and may not be in thermal equilibrium,
is driven to another QSS by the external field,
and the resulting QSS is not necessarily
in thermal equilibrium.
As a result, response to the external field
may differ from one obtained by statistical mechanics.
Indeed, in the ferromagnetic model
so-called Hamiltonian mean-field (HMF) model
\cite{inagaki-konishi-93,antoni-ruffo-95},
the critical exponents are obtained
as $\gamma_{-}=1/4$ \cite{ogawa-patelli-yamaguchi-14}
and $\delta=3/2$ \cite{ogawa-yamaguchi-14}
with the aid of a linear
\cite{patelli-gupta-nardini-ruffo-12,ogawa-yamaguchi-12}
and a nonlinear \cite{ogawa-yamaguchi-14} response theories
based on the Vlasov description respectively,
while statistical mechanics gives $\gamma_{-}=1$ and $\delta=3$.
Interestingly, with another exponent $\beta=1/2$,
the non-classical exponents satisfy
the classical scaling relation $\gamma_{-}=\beta(\delta-1)$,
and have universality on initial reference families
of QSSs in a wide class of 1D mean-field models
\cite{ogawa-yamaguchi-15}.

The universality is derived under the assumption
that the initial distribution functions depend on position and momentum
only through the one-particle Hamiltonian
with referring to the Jeans theorem \cite{jeans-15}.
Thus, the initial states are symmetric with respect to momentum.
The symmetric initial states are also used in studies on
nonequilibrium statistical mechanics
\cite{antoniazzi-07,chavanis-06,antoniazzi-07-prl,antoniazzi-07-prlb},
the core-halo description of QSSs \cite{pakter-levin-11},
nonequilibrium dynamics \cite{pakter-levin-13},
and correlation and diffusion \cite{yamaguchi-bouchet-dauxois-07}.
See also Refs.\cite{campa-dauxois-ruffo-09,levin-14}.

Nevertheless, asymmetric momentum distributions appear
in beam-plasma systems (see \cite{crawford-91,crawford-95,tsurutani-lakhina-97,karlicky-kasparova-09} for instance),
and are experimentally created in an ultracold plasma by optical pumping \cite{CBMPK-12}.
In the HMF model, such distributions are stationary even asymmetric,
and it is, therefore, natural to ask the response in the asymmetric case
for completing the response theory.
The main purpose of this article is to investigate the linear response 
against asymptotically constant external field
around spatially homogeneous but asymmetric distributions in the HMF model.
It is worth noting that, despite its simpleness, 
the model shares similar dynamics with the free-electron laser \cite{barre-04}
and an anisotropic Heisenberg model under classical spin dynamics
\cite{gupta-mukamel-11}.

The HMF model consists of plane rotators like the XY spins,
and the susceptibility
tensor in the HMF model is of size $2\times 2$
corresponding to the $x$- and $y$-directions of the rotators.
For symmetric homogeneous states,
the susceptibility tensor is directly diagonalized
and experiences a divergence at the critical point
of the second order phase transition,
which is dynamically interpreted as the stability threshold
of the homogeneous states
\cite{patelli-gupta-nardini-ruffo-12,ogawa-yamaguchi-12,ogawa-yamaguchi-15}.
We then ask the two questions for asymmetric
momentum distributions with zero-means:
Is the susceptibility tensor symmetric and diagonalizable ?
Does the response diverge at the stability threshold ?
We will answer these questions negatively.
The non-diagonalizable response tensor implies
that the external field for $x$-direction induces
the magnetization for $y$-direction,
and such a response is unavoidable even changing the coordinate.
Due to this non-diagonalizability,
the predicted stat is not stationary,
while the constant external field may drive the system
to a stationary state asymptotically.
In other words, the non-diagonalizability provides an example of
discrepancy between
the asymptotic states by the linear dynamics and the full Vlasov dynamics.
The non-divergence of response suggests
that $\gamma_{+}=0$ and $\delta=1$, and interestingly,
the scaling relation $\gamma_{+}=\beta(\delta-1)$ holds,
although $\beta$ might be not well defined
since the spatially inhomogeneous stationary states 
must be symmetric by the Jeans theorem \cite{jeans-15}.

This article is organized as follows.
The HMF model and the linear responses are reviewed
in Sec.\ref{sec:model-LRT}.
As an example of a family of asymmetric distributions,
we introduce the skew-normal distributions,
and investigate their stability in Sec.\ref{sec:skew-normal}.
Theoretical consequences are examined by direct numerical simulations
of the Vlasov equation in Sec.\ref{sec:numerics}.
We discuss on stationarity of the predicted state
in Sec.\ref{sec:stationarity-nonlinear}.
The last section \ref{sec:summary} is devoted to a summary and discussions.

\section{The Hamiltonian mean-field model and linear response theory}
\label{sec:model-LRT}

\subsection{The model}
The HMF model with the time-dependent external magnetic field
$\vec{h}=(h_{x}(t),h_{y}(t))$
is expressed by the Hamiltonian
\begin{equation}
    \begin{split}
        H_{N}(q,p,t)
        & = \sum_{j=1}^{N} \dfrac{p_{j}^{2}}{2}
        + \dfrac{1}{2N} \sum_{j,k=1}^{N} [ 1 - \cos (q_{j}-q_{k}) ] \\
        & - \sum_{j=1}^{N} [ h_{x}(t)\cos q_{j} + h_{y}(t)\sin q_{j} ].
    \end{split}
\end{equation}
The corresponding one-particle Hamiltonian is defined on the $\mu$ space,
which is $(-\pi,\pi]\times \mathbb{R}$, as
\begin{equation}
    \begin{split}
        \mathcal{H}[f](q,p,t)
        = \dfrac{p^{2}}{2}
        & - (M_{x}+h_{x}) \cos q
        - (M_{y}+h_{y}) \sin q
    \end{split}
\end{equation}
where the magnetization vector $(M_{x},M_{y})$ is defined by
\begin{equation}
    (M_{x},M_{y}) = \iint_{\mu} (\cos q,\sin q) f(q,p,t) {\rm d}q{\rm d}p.
\end{equation}
The one-particle distribution function $f$ is governed by the Vlasov equation
\begin{equation}
  \label{eq:Vlasov}
    \dfracp{f}{t}
    + \poisson{\mathcal{H}[f]}{f} = 0,
\end{equation}
with the Poisson bracket defined by
\begin{equation}
    \poisson{f}{g} = \dfracp{f}{p} \dfracp{g}{q} - \dfracp{f}{q} \dfracp{g}{p}.
\end{equation}
One can straightforwardly check that any spatially homogeneous states,
$f_{0}(p)$, are stationary if the external field $\vec{h}$ is absent.

We prepare a homogeneous stable stationary state
$f_{0}(p)$ for $t<0$,
and add a small external field $\vec{h}$ for $t>0$.
To avoid an artificial rotation,
we require the zero-mean for $f_{0}(p)$,
and consider an asymptotically constant external field accordingly.
For instance, we set
\begin{equation}
    \label{eq:external-field}
    \begin{pmatrix}
        h_{x}(t) \\ h_{y}(t)
    \end{pmatrix}
    = \Theta(t)
    \begin{pmatrix}
        h_{x} \\ h_{y}
    \end{pmatrix}
\end{equation}
using the Heaviside step function $\Theta(t)$,
and the external field drives the initial state $f_{0}$
to $f=f_{0}+f_{1}$ asymptotically.
Accordingly, the one-particle Hamiltonian $\mathcal{H}[f]$ changes from
$H_{0}$ to $H_{0}+H_{1}$, where
\begin{equation}
  \label{eq:H0-H1}
    H_{0} = \dfrac{p^{2}}{2} ,
    \quad
    H_{1} = -(M_{1,x}+h_{x})\cos q - (M_{1,y}+h_{y}) \sin q
\end{equation}
and
\begin{equation}
    M_{1,x} = \ave{\cos q}_{1},
    \quad
    M_{1,y} = \ave{\sin q}_{1}.
\end{equation}
We introduced the averages of an observable $B$
with respect to $f_{0}$ and $f_{1}$ as
\begin{equation}
    \ave{B}_{j} = \iint_{\mu} B(q,p) f_{j}(q,p) {\rm d}q{\rm d}p, \quad (j=0,1).  
\end{equation}

\subsection{Isothermal linear response}

It might be instructive to review the isothermal linear response
to compare it with the Vlasov linear response theory
which will be presented in the next subsection
\ref{sec:vlasov-linear-response}.

The thermal equilibrium states of the HMF model are describe
by the one-particle distribution functions of
\begin{equation}
    f(q,p) = \dfrac{e^{-\beta (H_{0}+H_{1})}}
    {\iint_{\mu} e^{-\beta (H_{0}+H_{1})}{\rm d}q{\rm d}p}.
\end{equation}
Hereafter $\beta$ represents not one of the critical exponents
mentioned in Sec.\ref{sec:introduction}, but the inverse temperature.
Expanding $f$ into the power series of $H_{1}$
and picking up to the linear order, we have
\begin{equation}
    \ave{B}_{1} = - \beta \left[ \ave{BH_{1}}_{0} - \ave{B}_{0} \ave{H_{1}}_{0}
    \right].
\end{equation}
Substituting $\cos q$ and $\sin q$ into $B$,
we have the matrix formula
\begin{equation}
    \label{eq:isothermal-matrix}
    \begin{pmatrix}
        M_{1,x} \\ M_{1,y}
    \end{pmatrix}
    = 
    \begin{pmatrix}
        C_{xx} & C_{xy} \\
        C_{yx} & C_{yy} \\
    \end{pmatrix}
    \left[
      \begin{pmatrix}
          M_{1,x} \\ M_{1,y}
      \end{pmatrix}
      + 
      \begin{pmatrix}
          h_{x} \\ h_{y}
      \end{pmatrix}
    \right],
\end{equation}
where the correlation matrix $C=(C_{\nu\sigma})~(\nu,\sigma\in\{x,y\})$ is defined by
\begin{equation}
    \label{eq:correlation-matrix}
    C = \beta
    \begin{pmatrix}
        \ave{\cos q\cos q}_{0} & \ave{\cos q\sin q}_{0} \\
        \ave{\sin q\cos q}_{0} & \ave{\sin q\sin q}_{0}
    \end{pmatrix}.
\end{equation}
Thus, the formal solution is
\begin{equation}
    \begin{pmatrix}
        M_{1,x} \\ M_{1,y}
    \end{pmatrix}
    = [ 1 - C ]^{-1} C
    \begin{pmatrix}
        h_{x} \\ h_{y}
    \end{pmatrix},
\end{equation}
and the susceptibility tensor $\chi=(\chi_{\nu\sigma})~(\nu,\sigma\in\{x,y\})$
defined by $\vec{M}=\chi\vec{h}$ in the limit $||\vec{h}||\to 0$ is
\begin{equation}
  \label{eq:isothermal-susceptibility-formal}
    \chi
    = [ 1-C ]^{-1} C.
\end{equation}
Divergence of $\chi$ appears at the critical point
satisfying $\det(1-C)=0$.

It is easy to show that the correlation matrix is now expressed by
$C=(\beta/2)I_{2}$, where $I_{2}$ is the $2\times 2$ unit matrix.
The susceptibility tensor is hence diagonalized and the diagonal elements are
\begin{equation}
    \label{eq:isothermal-susceptibility}
    \chi_{xx} = \chi_{yy} = \dfrac{\beta/2}{1-\beta/2}
    = \dfrac{T_{\rm c}}{T-T_{\rm c}}
\end{equation}
with the critical temperature $T_{\rm c}=1/2$
of the second order phase transition \cite{antoni-ruffo-95}.
The vanishing off-diagonal elements
come from spatial homogeneity of $f_{0}(p)$,
and symmetry of $f_{0}(p)$ is not necessary.

\subsection{Vlasov linear response}
\label{sec:vlasov-linear-response}

The nonlinear response theory \cite{ogawa-yamaguchi-14}
includes the linear response theory \cite{patelli-gupta-nardini-ruffo-12,ogawa-yamaguchi-12} for symmetric $f_{0}(p)$
and provides a simple expression of the linear response \cite{ogawa-yamaguchi-15},
but asymmetric $f_{0}(p)$ is out of range.
Thus, we revisit the linear response theory.

We introduce the Laplace transform defined by
\begin{equation}
  \label{eq:Laplace-trans}
  \widehat{u}(\omega) = \int_{0}^{\infty} u(t) e^{i\omega t} {\rm d}t.
\end{equation}
The linear response theory gives the Laplace transform
of $(M_{1,x}(t),M_{1,y}(t))$, denoted by $(\widehat{M}_{1,x}(\omega),\widehat{M}_{1,y}(\omega))$, as
\begin{equation}
    \label{eq:linear-response-HMF}
    \begin{pmatrix}
        \widehat{M}_{1,x}(\omega) \\
        \widehat{M}_{1,y}(\omega) \\
    \end{pmatrix}
    =
    [ 1 - F(\omega) ]^{-1} F(\omega)
      \begin{pmatrix}
        \widehat{h}_{x}(\omega) \\
        \widehat{h}_{y}(\omega) \\
    \end{pmatrix},
\end{equation}
where the elements of matrix $F=(F_{\nu\sigma})$ are
\begin{equation}
    \label{eq:F-elements}
    \begin{split}
        F_{xx}(\omega)
        & = \dfrac{-\pi}{2} \int_{L}
        \left( \dfrac{1}{p-\omega} + \dfrac{1}{p+\omega} \right) f'_{0}(p) {\rm d}p \\
        F_{xy}(\omega)
        & = \dfrac{-\pi}{2i} \int_{L}
        \left( \dfrac{1}{p-\omega} - \dfrac{1}{p+\omega} \right) f'_{0}(p) {\rm d}p \\
        F_{yx}(\omega)
        & = - F_{xy}(\omega) \\
        F_{yy}(\omega)
        & = F_{xx}(\omega).
    \end{split}
\end{equation}
See the Appendix \ref{sec:appendixA} for derivations.
The integral contour $L$ is the real $p$ axis for ${\rm Im}(\omega)>0$,
but is continuously modified for ${\rm Im}(\omega)\leq 0$
to avoid the poles at $p=\pm\omega$
by following the Landau's procedure \cite{landau-46}.

Temporal evolution of $(M_{1,x},M_{1,y})$ is determined
by performing the inverse Laplace transform,
which picks up singularities of its Laplace transform \eqref{eq:linear-response-HMF}.
For instance, a pole at $\omega_{\rm L}$ gives a term having
$\exp(-i\omega_{\rm L}t)$,
which implies the Landau damping for ${\rm Im}(\omega_{\rm L})<0$.
Assuming that the reference $f_{0}(p)$ is stable,
we have no singularities on the upper half $\omega$ plane.
Existence of singularities on the real axis of $\omega$
is accidental for $[1-F(\omega)]^{-1}F(\omega)$,
and we omit it.
Then, the main singularity comes from the Heaviside step function
of the external field \eqref{eq:external-field},
whose Laplace transform is
\begin{equation}
    \begin{pmatrix}
        \widehat{h}_{x}(\omega) \\
        \widehat{h}_{y}(\omega)
    \end{pmatrix}
    = \dfrac{-1}{i\omega}
    \begin{pmatrix}
        h_{x} \\ h_{y}
    \end{pmatrix}.
\end{equation}
Asymptotic values of $M_{1,x}$ and $M_{1,y}$ are,
therefore, obtained by picking up the pole at $\omega=0$
\cite{ogawa-yamaguchi-12}, and
\begin{equation}
    \begin{pmatrix}
        M_{1,x}(t) \\
        M_{1,y}(t)
    \end{pmatrix}
    \to
    \chi
    \begin{pmatrix}
        h_{x} \\ h_{y}
    \end{pmatrix}
    \quad
    (t\to\infty),
\end{equation}
where the susceptibility tensor $\chi=(\chi_{\nu\sigma})$ is
written in a similar form with \eqref{eq:isothermal-susceptibility-formal}
as
\begin{equation}
  \label{eq:vlasov-susceptibility-formal}
  \chi = [1-F(0)]^{-1}F(0).
\end{equation}

Let us rewrite the above Vlasov susceptibility $\chi$
by using the dispersion function
\begin{equation}
  \label{eq:dispersion-function}
  D(\omega) = 1 + \pi \int_{L} \dfrac{f'_{0}(p)}{p-\omega} {\rm d}p,
  \quad \omega\in\mathbb{C}.
\end{equation}
In the following we consider real $\omega$ which gives
\begin{equation}
  \label{eq:dispersion-function-real}
  D(\omega) = 1 + \pi~ {\rm PV}\int_{-\infty}^{\infty} \dfrac{f'_{0}(p)}{p-\omega} {\rm d}p
  + i\pi^{2} f'_{0}(\omega),
  \quad \omega\in\mathbb{R},
\end{equation}
where PV represents the principal value.
The dispersion function rewrites the susceptibility as
\begin{equation}
    \label{eq:vlasov-susceptibility}
    \chi = \dfrac{1}{|D(0)|^{2}}
    \begin{pmatrix}
        {\rm Re}(D(0))-|D(0)|^{2} & - {\rm Im}(D(0)) \\
        {\rm Im}(D(0)) &  {\rm Re}(D(0))-|D(0)|^{2}
    \end{pmatrix}.
\end{equation}

When $f_{0}(p)$ is symmetric and hence $f'_{0}(0)=0$,
implying ${\rm Im}(D(0))=0$ accordingly,
the susceptibility tensor $\chi$ is
diagonal, and the diagonal elements are
\begin{equation}
    \chi_{xx} = \chi_{yy} = \dfrac{1-D(0)}{D(0)}
\end{equation}
as reported in Refs.\cite{patelli-gupta-nardini-ruffo-12,ogawa-yamaguchi-12}.
The susceptibility, therefore, diverges at the point $D(0)=0$
corresponding to the stability threshold
\cite{inagaki-konishi-93,chavanis-delfini-09}.
On the other hand, when $f'_{0}(0)\neq 0$,
the imaginary part of $D(0)$ does not vanish and hence
the susceptibility tensor \eqref{eq:vlasov-susceptibility}
enjoys two interesting features:
(i) The tensor is neither symmetric nor diagonalizable
by the real coordinate transformation,
since the eigenvalues are not real.
(ii) No divergence appears even at the stability threshold,
since $|D(0)|^{2}>0$.
We note that, for homogeneous symmetric distributions,
$D(0)>0$ is the stability criterion
and hence the divergence appears at the stability threshold.
However, $D(0)>0$ is no more the stability criterion
for the asymmetric case.
A stability criterion for the asymmetric case
will be introduced in Sec.\ref{eq:nyquist-stability}.

\section{Skew-normal distribution and stability}
\label{sec:skew-normal}

\subsection{Skew-normal distribution}
We introduce the skew-normal distribution
for examining the linear response theory
and confirming the two features
mentioned in Sec.\ref{sec:vlasov-linear-response}.
Advantages of the skew-normal distribution are
that it has the single peak which makes the stability criterion simpler,
and that the analytically obtained mean value helps
to set the total momentum zero.

The density of skew-normal distribution is defined by
\begin{equation}
    f_{\rm SN}(x; \lambda, \mu, \sigma)
    = \dfrac{2}{\sigma} \phi\left( \dfrac{x-\mu}{\sigma} \right)
    \Phi\left( \lambda \dfrac{x-\mu}{\sigma} \right),
\end{equation}
where
\begin{equation}
    \phi(x) = \dfrac{1}{\sqrt{2\pi}} e^{-x^{2}/2}
\end{equation}
and
\begin{equation}
    \Phi(x) = \int_{-\infty}^{x} \phi(t) {\rm d}t
    = \dfrac{1}{2} \left[ 1 + {\rm erf}\left( \dfrac{x}{\sqrt{2}} \right) \right].
\end{equation}
The parameter $\lambda$ represents the skewness,
and $\lambda=0$ results to the normal distribution.
The mean value is
\begin{equation}
  \int_{-\infty}^{\infty} x f_{\rm SN} {\rm d}x
  = \mu + \sigma \delta \sqrt{\dfrac{2}{\pi}},
    \quad
    \delta = \dfrac{\lambda}{\sqrt{1+\lambda^{2}}}.
\end{equation}

\begin{figure}
    \centering
    \includegraphics[width=8cm]{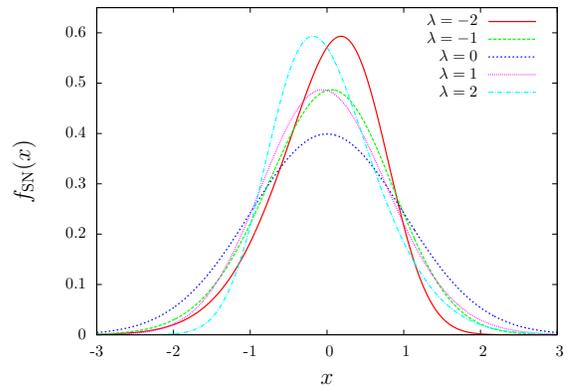}
    \caption{(color online)
      Skew-normal distributions with zero means
      and $\sigma=1$.  $\lambda=-2,-1,0,1$ and $2$,
      whose maximum points are from right to left.
      $f_{0}'(0)$ is positive (resp. negative)
      for negative (resp. positive) $\lambda$.}
    \label{fig:skew-normal}
\end{figure}

We test the homogeneous stationary states of the form
\begin{equation}
    f_{0}(p; \lambda, \mu, \sigma) = \dfrac{1}{2\pi} f_{\rm SN}(p; \lambda, \mu,\sigma),
\end{equation}
which is normalized as $\iint_{\mu}f_{0}{\rm d}q{\rm d}p=1$.
To set the total momentum zero, we put
\begin{equation}
    \mu = - \sigma\delta \sqrt{\dfrac{2}{\pi}}.
\end{equation}
Hereafter we fix the parameter $\sigma$ as $\sigma=1$.
Then, the unique free parameter is the skewness $\lambda$,
and the distribution is simply denoted by $f_{0}(p;\lambda)$.
Let $p=\eta$ be the unique extreme point (the maximum point)
depending on $\lambda$.
Some examples of the skew-normal distribution functions are exhibited
in Fig.\ref{fig:skew-normal}.

\subsection{Nyquist method of stability}
\label{eq:nyquist-stability}
For symmetric distributions $f_{0}(p)$,
the formal stability criterion has been established \cite{yamaguchi-04} as
\begin{equation}
    f_{0}(p) \text{ is formally stable}
    \quad \Longleftrightarrow \quad
    D(0) > 0,
\end{equation}
where $D$ is the dispersion function \eqref{eq:dispersion-function-real}.
To obtain the formal stability, $f_{0}(p)$ is assumed
as a function of one-particle Hamiltonian,
and hence we can not use this criterion for the skew-normal distributions.
Instead, we use the Nyquist method \cite{nyquist-32,nicholson-92},
which was applied to asymmetric double-peak distributions
in the HMF model \cite{chavanis-delfini-09}.

In our setting, the Nyquist method provides the stability criterion as
\begin{equation}
    \label{eq:penrose}
    \begin{split}
        & f_{0}(p;\lambda) \text{ has an exponentially growing mode} \\
        & \Longleftrightarrow \quad
        D(\eta)<0 
    \end{split}
\end{equation}
where $D(\omega)$ is the dispersion function \eqref{eq:dispersion-function-real} and is real at $\omega=\eta$.
See the Appendix \ref{sec:nyquist-method} for details.
The function $D(\eta)$ can be rewritten as
\begin{equation}
    \label{eq:Deta}
    D(\eta) = 1 + \pi \int_{-\infty}^{\infty}
    \dfrac{f_{0}(p;\lambda)-f_{0}(\eta;\lambda)}{(p-\eta)^{2}} {\rm d}p,
\end{equation}
by performing the integration by parts
and remembering $f'_{0}(\eta;\lambda)=0$ \cite{penrose-60}.
The Taylor expansion says that the numerator of the integrand
starts from the quadratic term, $(p-\eta)^{2}$,
and hence no singularity appears in the integrand.
A rigorous treatment of the above Penrose criterion
is found in Ref.\cite{faou-rousset-14}.

\begin{figure}
    \centering
    \includegraphics[width=8cm]{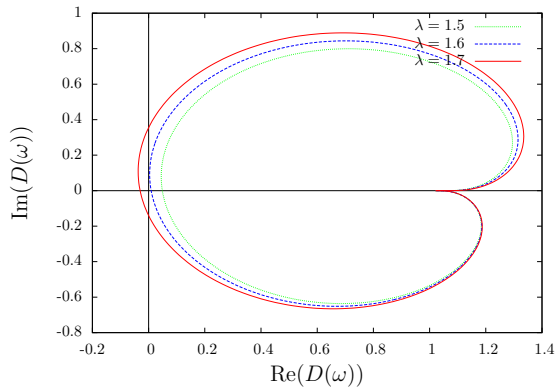}
    \caption{(color online)
      Nyquist diagrams for the skew-normal distributions $f_{0}(p;\lambda)$
      with $\lambda=1.5$ (green dotted), $1.6$ (blue dashed)
      and $1.7$ (red solid).
      Each curve is the mapped real $\omega$ axis by $D$,
      which intersects with the real $D(\omega)$ axis at $\omega=\eta$,
      the unique extreme point.
      Inside of the curve corresponds to the upper half $\omega$ plane.}
    \label{fig:Nyquist}
\end{figure}

The stability criterion \eqref{eq:penrose} is graphically
presented in Fig.\ref{fig:Nyquist}.
The mapped real $\omega$ axis by $D$
intersects with the real $D(\omega)$ axis
at $\omega=\eta$ only, since ${\rm Im}(D(\omega))$ 
vanishes at the unique extreme point.
Consequently, we can say that the state $f_{0}(p;\lambda)$ is unstable
iff the mapped real $\omega$ axis by $D$ crosses
with the negative real axis on the complex $D(\omega)$ plane.
Observing Fig.\ref{fig:Nyquist},
the stability threshold of the skew-normal distributions,
denoted by $\lambda_{\rm th}$,
must be in the interval $1.6<\lambda_{\rm th}<1.7$.
From symmetry with respect to $\lambda$,
we have another threshold $-\lambda_{\rm th}$,
and $f_{0}(p;\lambda)$ is stable for 
$-\lambda_{\rm th}<\lambda<\lambda_{\rm th}$.

\begin{figure}
    \centering
    \includegraphics[width=8cm]{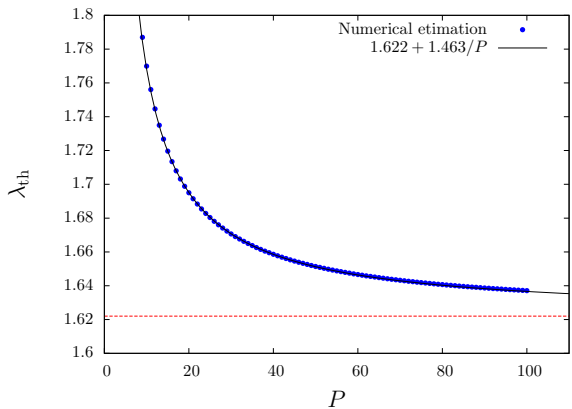}
    \caption{(color online)
      Numerical estimation of threshold $\lambda_{\rm th}$
      with varying cut-off $P$ (blue circles).
      The black solid curve is the fitting by the least squares method
      in the interval $[10,100]$ of $P$,
      and the red horizontal dashed line is the estimated level
      of $\lambda_{\rm th}=1.622$.}
    \label{fig:Penrose}
\end{figure}

The stability threshold can be estimated by precise numerical computations.
The integral in Eq.\eqref{eq:Deta} is in an infinite interval,
and is impossible to perform exactly in numerics.
To estimate the infinite interval integration,
we introduce the cut-off $P$ as 
\begin{equation}
    \label{eq:DetaP}
    D_{P}(\eta) = 1 + \pi \int_{-P}^{P}
    \dfrac{f_{0}(p;\lambda)-f_{0}(\eta;\lambda)}{(p-\eta)^{2}} {\rm d}p,
\end{equation}
and observe $P$-dependence of $\lambda_{\rm th}$.
Estimated threshold with varying $P$ is reported in Fig.\ref{fig:Penrose},
and is fitted by $1.622+1.463/P$,
where the fitting curve is obtained by the least squares method.
We hence conclude that the threshold is $\lambda_{\rm th}\simeq 1.622$
in the limit $P\to\infty$.

\section{Numerical tests}
\label{sec:numerics}

We use the semi-Lagrangian code \cite{debuyl-10}
with the time slice $\Delta t=0.05$.
The $\mu$ space, the $(q,p)$ plane, is truncated to $(-\pi,\pi]\times [-4,4]$,
and is divided into $G\times G$ grid points.
We call $G$ the grid size.
The magnetization is zero for the reference homogeneous state
$f_{0}(p;\lambda)$,
and therefore, we simply denote the response magnetization
as $(M_{x},M_{y})$ instead of $(M_{1,x},M_{1,y})$.

It might be worth remarking that the truncation at $|p|=4$
does not conflict with the estimation of $\lambda_{\rm th}$
reported in Fig.\ref{fig:Penrose}, which requires a larger cut-off.
The reference state $f_{0}$ rapidly decreases as the Gaussian,
while the integrand in \eqref{eq:DetaP} slowly decreases as $p^{-2}$
in the large $|p|$ due to existence of the constant $f_{0}(\eta)$.

\subsection{Stability threshold and unstable branch}
\begin{figure}
    \centering
    \includegraphics[width=8cm]{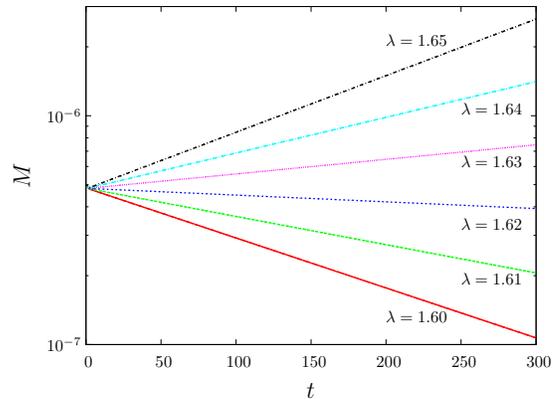}
    \caption{(color online)
      Initial temporal evolutions of $M$
      for the perturbed initial state $f_{\epsilon}(q,p;\lambda)$,
      \eqref{eq:perturbed}, with $\epsilon=10^{-6}$ and
      $\lambda=1.60, 1.61, 1.62, 1.63, 1.64$ and $1.65$
      from bottom to top. The grid size is $G=512$.
      The vertical axis is in logarithmic scale.
      The stability threshold is in the interval $1.62<\lambda_{\rm th}<1.63$,
      and is consistent with the estimated value $\lambda_{\rm th}\simeq 1.622$.}
    \label{fig:Stability}
\end{figure}

The obtained stability threshold is directly examined
by computing temporal evolution of a perturbed state.
We prepare the perturbed initial state as
\begin{equation}
    \label{eq:perturbed}
    f_{\epsilon}(q,p;\lambda) = f_{0}(p;\lambda) (1 + \epsilon \cos q),
\end{equation}
and use $\epsilon=10^{-6}$. Temporal evolution of $M=(M_{x}^{2}+M_{y}^{2})^{1/2}$
is shown in Fig.\ref{fig:Stability},
and the computed threshold $\lambda_{\rm th}$ is successfully confirmed.

\begin{figure}
    \centering
    \includegraphics[width=8cm]{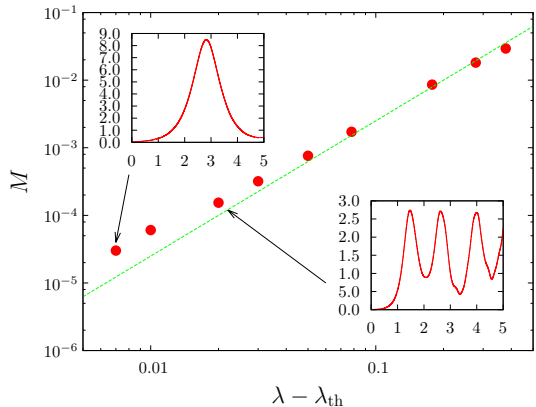}
    \caption{(color online)
      Time averaged $M$ as a function of $\lambda-\lambda_{\rm th}$
      for $f_{\epsilon}(q,p;\lambda)$ \eqref{eq:perturbed}
      with $\epsilon=10^{-6}$.
      The time window for averages is $[1000,5000]$.
      The grid size is $G=512$.
      The green straight line represents $M=(\lambda-\lambda_{\rm th})^{2}/4$
      for guide of eyes.
      The insets represent temporal evolutions of $M$ for the marked points.
      The horizontal axis represents the scaled time $t/1000$,
      and the vertical axes $10^{5}M$ and $10^{4}M$
      for the upper-left and the lower-right insets respectively.}
    \label{fig:ResponseUnstable}
\end{figure}

When the initial state is symmetric with respect to $p$,
the nonlinear response theory \cite{ogawa-yamaguchi-14} predicts that
$M$ will be proportional to $(\lambda-\lambda_{\rm th})^{2}$
in the unstable branch.
Numerical simulations captured oscillations of $M$
around the predicted levels 
and the period tends to increase as the initial state approaches
to the stability threshold \cite{ogawa-yamaguchi-14}.
Even the present asymmetric case, the scaling, oscillations
and a similar tendency of periods are observed
as reported in Fig.\ref{fig:ResponseUnstable}.

\subsection{Linear responses}

We come back to the unperturbed initial distribution $f_{0}(p;\lambda)$,
and add the external field \eqref{eq:external-field}.
From symmetry of the system
we set $(h_{x},h_{y})=(h,0)$ without loss of generality.

\begin{figure}
    \centering
    \includegraphics[width=8cm]{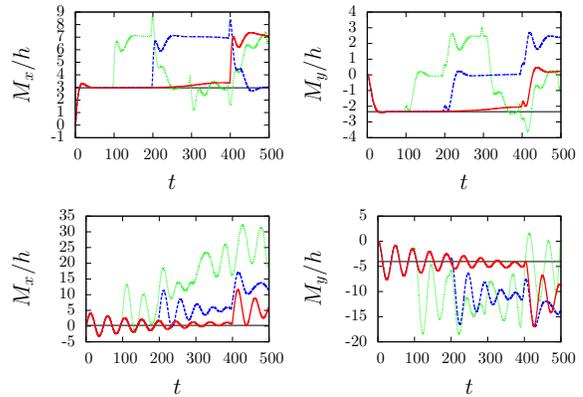}
    \caption{(color online)
      Normalized responses $M_{x}/h$ (left) and $M_{y}/h$ (right)
      with $h=10^{-5}$.
      $\lambda=1.2$ (upper) and $1.6$ (lower).
      The grid sizes are $G=128$ (green dotted),
      $256$ (blue dashed) and $512$ (red solid).
      The black horizontal lines are theoretical predictions:
      $M_{x}/h=2.972$ and $M_{y}/h=-2.344$ for $\lambda=1.2$,
      and $M_{x}/h=0.2353$ and $M_{y}/h=-4.042$ for $\lambda=1.6$.}
    \label{fig:Response}
\end{figure}

In order to examine the linear response theory,
we set $h=10^{-5}$ to be small enough.
The normalized responses $M_{x}/h$ and $M_{y}/h$,
which are susceptibilities in the limit $h\to 0$,
are reported in Fig.\ref{fig:Response}
for stable states of $\lambda=1.2$ and $1.6$.

The theoretically predicted levels of responses are in good agreements
with the numerical experiments in initial time intervals.
The life time of the agreements gets longer
as the grid size $G$ increases,
and is, roughly speaking, proportional to $G$.
We may therefore conclude that the theoretically predicted response tensor
is valid for a long time
and that the non-zero off-diagonal response is observable
if we use a fine grid.

\begin{figure}
    \centering
    \includegraphics[width=8cm]{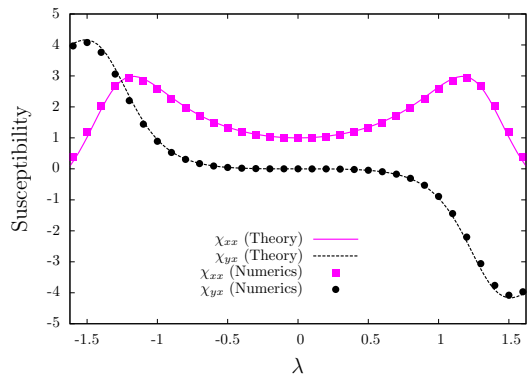}
    \caption{(color online)
      Elements of susceptibility tensor as a function of the skewness $\lambda$.
      Lines are from theory.
      Points are from numerics with the grid size $G=512$,
      and $M_{x}$ and $M_{y}$ are computed as averages
      over the time window $[0,200]$.
      Diagonal element $\chi_{xx}$ (magenta solid/squares),
      and off-diagonal element $\chi_{yx}$ (black dashed/circles).
      The region of $\lambda$ is restricted in the stable interval.}
    \label{fig:ResponseTheory}
\end{figure}

For the whole stable region of $\lambda$,
the theory is compared with numerical results in Fig.\ref{fig:ResponseTheory}.
We remark that the state with $\lambda=0$ is the thermal equilibrium state
of temperature $T=1$, and the normalized response $M_{x}/h$
coincides with the previously computed Vlasov linear response
$T_{\rm c}/(T-T_{\rm c})=1$
\cite{patelli-gupta-nardini-ruffo-12,ogawa-yamaguchi-12},
which is also coincides with isothermal linear response
\eqref{eq:isothermal-susceptibility}.
We stress that, as stated in the end of Sec.\ref{sec:vlasov-linear-response},
no divergence is observed at the stability threshold,
which are the left and right boundaries of the figure.
Another remark is that strength of response
$(M_{x}^{2}+M_{y}^{2})^{1/2}/h$ for $\lambda\neq 0$
is greater than the symmetric case, $\lambda=0$.

One possible explanation for the sign of $\chi_{yx}$ is as follows.
We may concentrate for $\lambda>0$ without loss of generality.
In this case the negative part of $f_{0}(p;\lambda)$ is larger than
the positive part around $p=0$,
and hence the small cluster being around $p=0$
induced by the external field locally has negative total momentum.
Consequently, the magnetization vector turns to the negative direction of $q$
by the external field.

\subsection{Dependence on external magnetic field}

The present non-diagonalizable susceptibility tensor
comes from non-zero $f'_{0}(0;\lambda)$,
which implies that the maximum point $\eta$ differs from the origin.
Thus, we expect that asymmetric characters of the linear response
tend to be hidden if the characteristic scale of $p$-axis,
width of the separatrix, is larger than the maximum point $p=\eta$,
since the local total momentum in the separatrix
approaches to zero.

For the magnetization $(M_{x},M_{y})$ and the external field $(h,0)$,
the separatrix reaches to $|p|=2\sqrt{||{\vec{M}}||+h}$.
The magnetization is induced by the external field, and we have
\begin{equation}
    ||\vec{M}|| = h \sqrt{(\chi_{xx})^{2}+(\chi_{yx})^{2}}.
\end{equation}
Then, we may expect that the asymmetric characters
appear for small $h$ satisfying
\begin{equation}
    h < h_{\rm th}, 
    \quad
    h_{\rm th} = \dfrac{\eta^{2}}{4 \left[ (\chi_{xx})^{2}+(\chi_{yx})^{2} + 1 \right]}.
\end{equation}

\begin{figure}
    \centering
    \includegraphics[width=8cm]{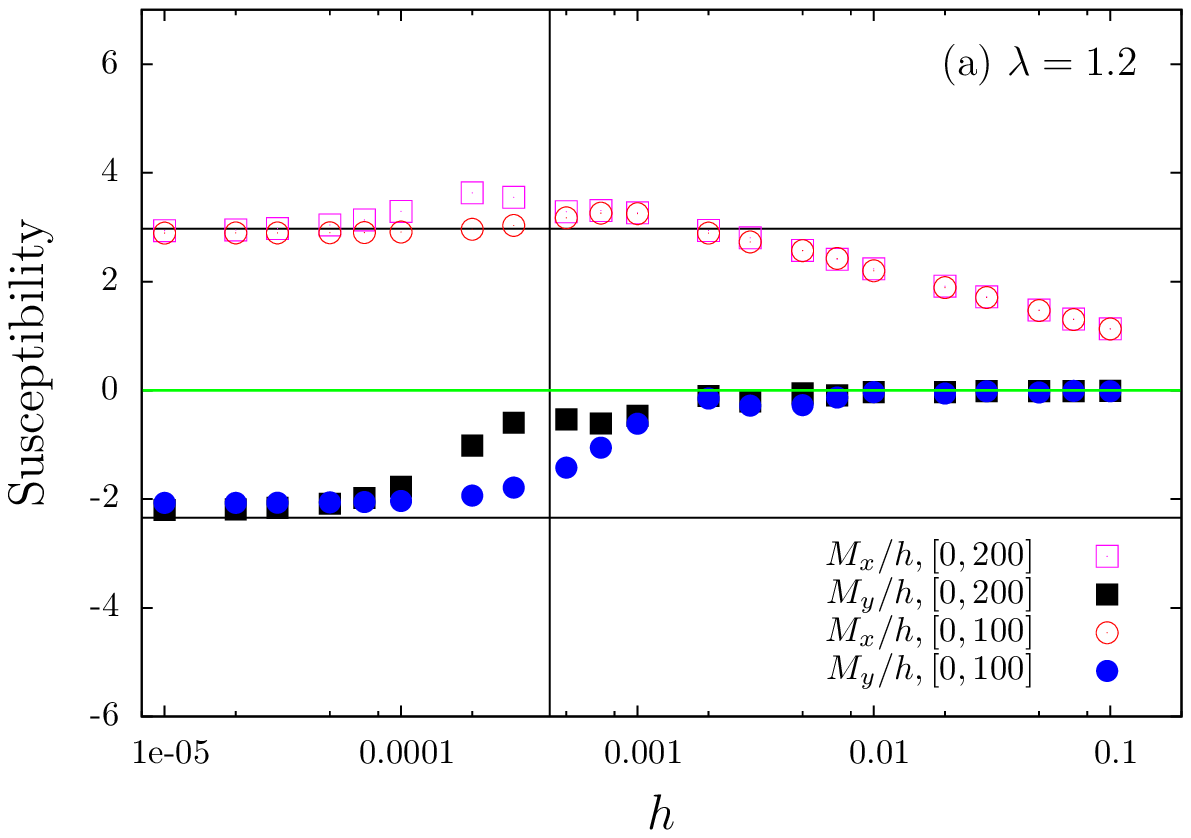}
    \includegraphics[width=8cm]{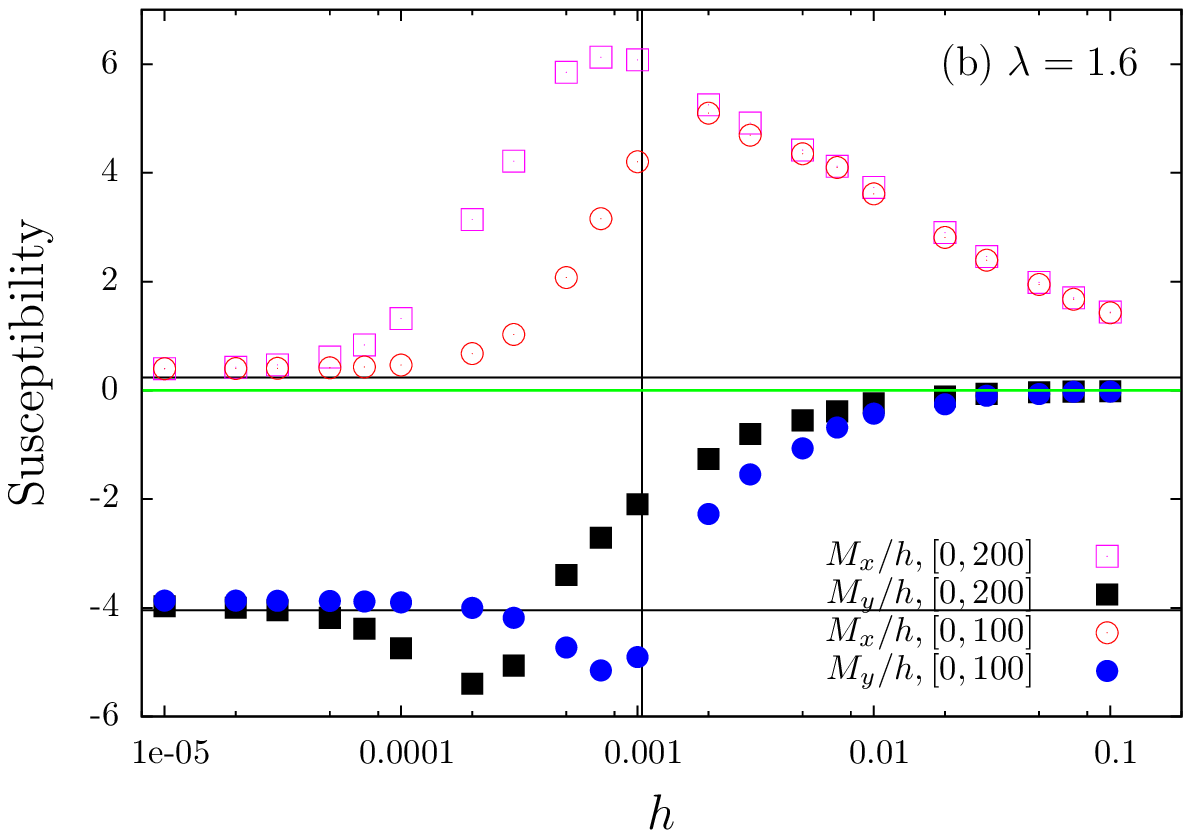}
    \caption{(color online)
      $h$ dependence of susceptibilities for
      (a) $\lambda=1.2$ and (b) $\lambda=1.6$.
      Open symbols are for $M_{x}/h$,
      and filled symbols for $M_{y}/h$,
      which are averaged over the time windows $[0,200]$ (squares)
      or $[0,100]$ (circles).
      The vertical black lines represent $h_{\rm th}$,
      and horizontal black lines the linear response levels.
      The horizontal green lines are the zero level.
      The grid size is $G=512$.}
    \label{fig:hdep}
\end{figure}

We report $h$ dependence of susceptibilities in Fig.\ref{fig:hdep}
for $\lambda=1.2$ and $1.6$.
The normalized responses, $M_{x}/h$ and $M_{y}/h$,
approaches to the theoretically predicted levels in $h<h_{\rm h}$,
while the off-diagonal response, $M_{y}/h$, goes to zero for larger $h$.

\section{Stationarity and nonlinear effects}
\label{sec:stationarity-nonlinear}

Let us discuss a possible scenario of temporal evolution
with off-diagonal response.
First of all, we show the fact that the predicted state with non-zero $M_{y}$
is not stationary by stating that $\vec{M}$ and $\vec{h}$ must be parallel
in a stationary state.

Jeans theorem \cite{jeans-15,binney-tremaine-08}
states that an inhomogeneous distribution function is a stationary solution
of the Vlasov equation if and only if it depends on $(q,p)$ only through
integrals of the one-particle Hamiltonian system.
The responded state has non-zero $(M_{x},M_{y})$
and the integral is the Hamiltonian
\begin{equation}
    \mathcal{H} = p^{2}/2 - \tilde{M}\cos(q-\alpha),
\end{equation}
where
\begin{equation}
    \tilde{M} = \sqrt{(M_{x}+h_{x})^{2}+(M_{y}+h_{y})^{2}},
    \quad
    \tan\alpha = \dfrac{M_{y}+h_{y}}{M_{x}+h_{x}}.
\end{equation}
Then, for a stationary state $f(\mathcal{H}(q,p))$,
we have the vanishing integral of
\begin{equation}
    0 = \iint_{\mu} \sin (q-\alpha) f(\mathcal{H}(q,p)) {\rm d}q{\rm d}p
    = M_{y}\cos\alpha - M_{x}\sin\alpha,
\end{equation}
since the integrand of the middle term is odd with respect to $q-\alpha$.
This equality and the definition of $\alpha$ imply
\begin{equation}
    \dfrac{M_{y}+h_{y}}{M_{x}+h_{x}} = \dfrac{M_{y}}{M_{x}},
\end{equation}
and we conclude $\vec{M}$ and $\vec{h}$ are parallel.

As a result, the state predicted by the linear response theory
is not a stationary state, and hence the system does not keep
the predicted state as observed in Fig.\ref{fig:Response}.
We can point out a similarity of the present phenomenon
with the nonlinear trapping \cite{oneil-65}.
If the Landau damping time scale is longer
than the so-called trapping time scale,
then the exponential Landau damping stops
and a cluster is formed by nonlinear effects \cite{barre-yamaguchi-09}.
In other words,
the state experiences the linear Landau damping in an early time interval,
but stops to damp by the nonlinear effects.
Similarly, the state predicted by the linear response theory appears
in a short time interval, and then disappears.
We conjecture that the disappearance comes from nonlinearity
of the full Vlasov equation.

\section{Summary and discussions}
\label{sec:summary}

We investigated the response tensor against an asymptotically constant external field
for spatially homogeneous but asymmetric momentum distributions
with zero-means by using the linear response theory.
The theory predicts two interesting characters of the susceptibility tensor:
One is non-diagonalizablility,
and the other is non-divergence even at the stability threshold.
The first character implies that
the external field added to the $x$-direction
induces the magnetization to the $y$-direction
even in the simple HMF model.
The off-diagonal response is not mysterious in our setting,
since anisotropy is included in asymmetry of momentum distributions.
For realizing the theoretical setting,
we introduced a family of skew-normal distributions.
After studying stability of the family by the Nyquist method,
all the theoretical consequences are successfully confirmed
by direct numerical simulations of the Vlasov equation.
We stress that the crucial condition for the two characters
is non-zero derivative of the reference state,
$f_{0}'(0)\neq 0$, which never happens for symmetric $f_{0}(p)$.
One physical example of $f_{0}'(0)\neq 0$ can be found in a beam-plasma system,
whose momentum distribution consists of, for instance,
a drifting Maxwellian for the beam and a Maxwellian for the plasma
\cite{crawford-91}.
In this example the non-zero derivative $f_{0}'(0)\neq 0$ is realized
both with and without shifting the distribution
to set the total momentum zero in general.
Studying distributions having two or many peaks is a future work.

The state reached by the linear response
is neither in thermal equilibrium nor in a stationary state,
since the off-diagonal response is not zero,
while the magnetization and the external field vectors must be parallel
in a stationary state.
The life time of such a state is finite,
but gets longer as the grid size becomes finer.
Thus, we may expect that the off-diagonal response
can be experimentally observed by using large enough number of particles.
However, non-stationarity may cause shortness of the life time
comparing with the symmetric case,
and revealing the time scale in which the linear response theory is valid
is remained as another future work.

Concerning to the above discussion,
we remark on validity of the linear response theory
to predict asymptotic stationary states.
We considered stable reference states, and added an external field small enough.
Nevertheless, the asymptotic stationary states cannot be predicted
by the linear response theory for asymmetric homogeneous initial states.
Analogy with the linear Landau damping might be interesting,
which is stopped by nonlinear effects.
Recently nonlinear equations for magnetization moments has been proposed
for homogeneous waterbag initial distributions
in the HMF model under an external field \cite{pakter-levin-13}.
An extension to non-waterbag states possibly helps to understand
the nonlinear effects and to solve the puzzle on the linear response theory.

In addition to the stable initial states,
perturbed unstable asymmetric initial states are also studied,
and similar features are numerically observed with the symmetric case
\cite{ogawa-yamaguchi-14}
in ordering and oscillations of magnetization
around the saturated states.
Apart from the macroscopic variable,
looking into difference in distribution functions is a remaining work.
For instance, the core-halo structure \cite{levin-14} has been observed
on the $\mu$ space for waterbag initial states \cite{pakter-levin-11},
but it is still unclear if the present asymmetric unstable states
also yield such structure in the saturated states.

In this article we focused on the asymptotically constant external field
corresponding to the zero total momentum,
but an oscillating external field of $\cos(\omega_{0}t)~(\omega_{0}\in\mathbb{R})$
is also available.
Laplace transform of the external field provides poles at $\omega=\pm\omega_{0}$,
and the denominator of susceptibility, $|D(0)|^{2}$,
is replaced with $D(\pm\omega_{0})\overline{D(\mp\omega_{0})}$
as shown in \eqref{eq:general-chi},
where $\overline{D(\omega_{0})}$ is the complex conjugate of $D(\omega_{0})$.
As a result, setting $\omega_{0}=\eta$ where $\eta$ is the maximum point
of momentum distribution,
the susceptibility diverges at the stability threshold,
which satisfies $D(\eta)=0$.
The symmetry is, therefore, not essential for the divergence of susceptibility.
Even in this case, the susceptibility tensor has non-zero off-diagonal
elements reflecting the asymmetry, see \eqref{eq:chixy-eta}.

\acknowledgements
The author thanks the anonymous referees for useful comments
to improve the manuscript.
He acknowledges the support of
JSPS KAKENHI Grant Number 23560069.

\appendix

\section{Derivation of Vlasov linear response}
\label{sec:appendixA}
Let $X_{0}$ be the Hamiltonian vector field
associated with the Hamiltonian $H_{0}$, \eqref{eq:H0-H1},
which is expressed as
\begin{equation}
    X_{0} = p \dfracp{}{q}.
\end{equation}
Linearizing the Vlasov equation \eqref{eq:Vlasov} around $f_{0}(p)$,
we have the formal solution of perturbation $f_{1}(q,p,t)$ as
\begin{equation}
    f_{1}(q,p,t) = - \int_{0}^{t} e^{-(t-s)X_{0}}
    \poisson{H_{1}(s)}{f_{0}} {\rm d}s
\end{equation}
for the initial condition $f_{1}(q,p,t=0)=0$.
The operator $\exp(tX_{0})$ acts on a function $u(q,p)$ as
\begin{equation}
    e^{tX_{0}} u(q,p) = u(\varphi_{0}^{t}(q,p)),
\end{equation}
where $\varphi_{0}^{t}$ is the Hamiltonian flow
associated with $H_{0}$ and hence $\varphi_{0}^{t}(q,p)=(q+pt,t)$
in our setting. We can prove the equality
\begin{equation}
    \iint_{\mu} v(q,p) u(\varphi_{0}^{-t}(q,p)) {\rm d}q{\rm d}p
    = \iint_{\mu} v(\varphi_{0}^{t}(q,p)) u(q,p) {\rm d}q{\rm d}p
\end{equation}
by changing variables $(q',p')=\varphi_{0}^{-t}(q,p)$
and using ${\rm d}q'{\rm d}p'={\rm d}q{\rm d}p$
from canonical property of $\varphi_{0}^{t}$.
Thus, we have
\begin{equation}
    \ave{B}_{1}(t) = -\iint_{\mu} {\rm d}q{\rm d}p \int_{0}^{t} B_{t-s}(q,p)
    \poisson{H_{1}(s)}{f_{0}}  {\rm d}s,
\end{equation}
where $B_{t}(q,p)=B(\varphi_{0}^{t}(q,p))$.
Performing the Laplace transform \eqref{eq:Laplace-trans},
we obtain
\begin{equation}
    \label{eq:linear-response}
    \widehat{\ave{B}_{1}}(\omega)
    = - \iint_{\mu} \widehat{B}_{\omega}(q,p)
    \poisson{\widehat{H}_{1}(q,\omega)}{f_{0}(p)} {\rm d}q{\rm d}p
\end{equation}
with
\begin{equation}
    \begin{split}
        \widehat{H}_{1}(q,\omega)
        & = - \left[ \widehat{M}_{1,x}(\omega) + \widehat{h}_{x}(\omega) \right]
        \cos q \\
        & - \left[ \widehat{M}_{1,y}(\omega) + \widehat{h}_{y}(\omega) \right]
        \sin q.
    \end{split}
\end{equation}

Substituting $B=\cos q$ and $B=\sin q$
into the linear response formula \eqref{eq:linear-response},
and using the Laplace transforms of
$\cos q_{t}=\cos(q+pt)$ and $\sin q_{t}=\sin(q+pt)$, which are
respectively
\begin{equation}
    \widehat{\cos q}_{\omega}
    = \dfrac{1}{2i} \left( \dfrac{e^{-iq}}{p-\omega}
      - \dfrac{e^{iq}}{p+\omega} \right)
\end{equation}
and
\begin{equation}
    \widehat{\sin q}_{\omega}
    = \dfrac{1}{2} \left( \dfrac{e^{-iq}}{p-\omega}
      + \dfrac{e^{iq}}{p+\omega} \right),
\end{equation}
we have the matrix form of
\begin{equation}
    \label{eq:linear-response-HMF-formal}
    \begin{pmatrix}
        \widehat{M}_{1,x}(\omega) \\
        \widehat{M}_{1,y}(\omega) \\
    \end{pmatrix}
    =
    \begin{pmatrix}
        F_{xx}(\omega) & F_{xy}(\omega) \\
        F_{yx}(\omega) & F_{yy}(\omega) \\
    \end{pmatrix}
    \left[
      \begin{pmatrix}
          \widehat{M}_{1,x}(\omega) \\
          \widehat{M}_{1,y}(\omega) \\
      \end{pmatrix}
      +
      \begin{pmatrix}
        \widehat{h}_{x}(\omega) \\
        \widehat{h}_{y}(\omega) \\
    \end{pmatrix}
    \right] .
\end{equation}
The elements of the matrix $F$ are exhibited in \eqref{eq:F-elements}.

To ensure convergence of the Laplace transform \eqref{eq:Laplace-trans},
the matrix $F(\omega)$ is defined in the upper half $\omega$ plane.
We analytically continue the domain into the whole complex $\omega$ plane \cite{landau-46},
and the resulting integral is written as
\begin{equation}
  \label{eq:continued-integral}
  \int_{L} \dfrac{f'_{0}(p)}{p\mp\omega} {\rm d}p
  = {\rm PV} \int_{-\infty}^{\infty} \dfrac{f'_{0}(p)}{p\mp\omega} {\rm d}p
  \pm S(\omega) i\pi f'_{0}(\pm\omega)
\end{equation}
where ${\rm PV}$ represents the principal value
and is the normal integral for $\omega\not\in\mathbb{R}$,
and the second term including
\begin{equation}
  S(\omega) = \left\{
    \begin{array}{ll}
      0, & {\rm Im}(\omega)>0 \\
      1, & {\rm Im}(\omega)=0 \\
      2, & {\rm Im}(\omega)<0 \\
    \end{array}
  \right.
\end{equation}
comes from the residues.

We remark that the linear response \eqref{eq:linear-response}
is rewritten as
\begin{equation}
    \label{eq:linear-response-parts}
    \widehat{\ave{B}_{1}}(\omega)
    = - \ave{ \poisson{ \widehat{B}_{\omega}(q,p) }{ \widehat{H}_{1}(q,\omega) } }_{0},
\end{equation}
if we perform integration by parts.
The expression \eqref{eq:linear-response-parts} gives
a similar form of the matrix $F$ with the correlation matrix $C$
\eqref{eq:correlation-matrix} as
\begin{equation}
    F(\omega) = 
    \begin{pmatrix}
        \ave{ \poisson{ \widehat{\cos q}_{\omega} }{\cos q } }_{0}
        & \ave{ \poisson{ \widehat{\cos q}_{\omega} }{\sin q } }_{0} \\
        \ave{ \poisson{ \widehat{\sin q}_{\omega} }{\cos q } }_{0} 
        & \ave{ \poisson{ \widehat{\sin q}_{\omega} }{\sin q } }_{0}
    \end{pmatrix} .
\end{equation}
The matrix $F$ coincides with the correlation matrix $C$ as
$F(\omega)=(\beta/2)I_{2}$
if $f_{0}(p)$ is the Maxwellian with the inverse temperature $\beta$.
Therefore, the Vlasov linear response coincides with the isothermal linear response
in thermal equilibrium of the homogeneous phase \cite{patelli-gupta-nardini-ruffo-12,ogawa-yamaguchi-12}.

In the text we concentrated on response to the external field with $\omega=0$,
but a general $\omega$ is also available.
The explicit form of the matrix $[1-F(\omega)]^{-1}F(\omega)$ is
\begin{equation}
  \label{eq:general-chi}
  [1-F(\omega)]^{-1}F(\omega)
  = \dfrac{1}{D(\omega)\overline{D(-\overline{\omega})}}
  \begin{pmatrix}
    G(\omega) & F_{xy}(\omega) \\
    - F_{xy}(\omega) & G(\omega)
  \end{pmatrix}
\end{equation}
where $\overline{\omega}$ is the complex conjugate of $\omega$ and
\begin{equation}
  G(\omega) = [1-F_{xx}(\omega)]F_{xx}(\omega)-[F_{xy}(\omega)]^{2}.
\end{equation}
In particular, the off-diagonal element is written by
\begin{equation}
  \begin{split}
    F_{xy}(\omega)
    = & - \dfrac{\pi}{2i} \left[ {\rm PV}\int_{-\infty}^{\infty} \dfrac{f'_{0}(p)}{p-\omega} {\rm d}p
      - {\rm PV}\int_{-\infty}^{\infty} \dfrac{f'_{0}(p)}{p+\omega} {\rm d}p \right] \\
    & - S(\omega) \dfrac{\pi^{2}}{2} \left[ f'_{0}(\omega) + f'_{0}(-\omega) \right]
  \end{split}
\end{equation}
and results to $-{\rm Im}(D(0))=-\pi^{2}f'_{0}(0)$ at $\omega=0$
as shown in the susceptibility \eqref{eq:vlasov-susceptibility}.
If we consider the oscillating external field of
$\cos(\omega_{0}t)~(\omega_{0}\in\mathbb{R})$,
the susceptibility becomes
\begin{equation}
  2\chi = [1-F(\omega_{0})]^{-1}F(\omega_{0})
  + [1-F(-\omega_{0})]^{-1}F(-\omega_{0}).
\end{equation}
Thus, for $\omega_{0}=\eta$,
where $\eta$ is the unique extreme point of $f_{0}(p)$,
the diagonal elements of susceptibility diverges
at the stability threshold satisfying $D(\eta)=0$.
Even in this case, the oscillating external field 
gives the non-zero off-diagonal element as
\begin{equation}
  \label{eq:chixy-eta}
  \chi_{xy}
  = \dfrac{-\pi^{2}f_{0}'(-\eta)/2}{
    \left(1+\pi{\rm PV}\int \dfrac{f_{0}'(p)}{p+\eta} dp \right)^{2}
    + \left( \pi^{2}f_{0}'(-\eta) \right)^{2} }.
\end{equation}

\section{Nyquist method}
\label{sec:nyquist-method}
To review the Nyquist method,
we restrict ourselves in single-peak distributions including
the skew-normal distributions.
Let us define the set $R=\{ D(\omega)\in\mathbb{C} ~|~ {\rm Im}(\omega)> 0\}$,
where $D(\omega)$ is the dispersion function \eqref{eq:dispersion-function}.
If this set $R$ includes the origin,
then there exists a root of the dispersion relation $D(\omega)$
on the upper half $\omega$ plane,
and the root corresponds to an exponential growing mode
from the definition of the Laplace transform \eqref{eq:Laplace-trans}.

To study the set $R$, we investigate the boundary
\begin{displaymath}
  \partial R=\{ D(\omega) \in \mathbb{C} ~|~ {\rm Im}(\omega)=0 \}.
\end{displaymath}
The boundary forms a closed curve,
since $D(\omega)\to 1$ as $\omega\to\pm\infty$.
In the limits of $\omega\to -\infty$ and $+\infty$,
the curve approaches to $1$
from the positive and the negative imaginary sides respectively,
since $f'_{0}(p)>0$ for $p<\eta$ and $f'_{0}(p)<0$ for $p>\eta$,
where $\eta$ is the maximum point of the single-peak distribution $f_{0}(p)$.
Then, the orientation implies that
the upper half $\omega$ plane is mapped onto the inside
of the closed curve.
The imaginary part of $D(\omega)$ is proportional to $f'_{0}(\omega)$
for $\omega$ real,
and vanishes if and only if $\omega$ coincides
with the unique extreme point $\eta$.
Thus, $D(\eta)$ is real and $D(\eta)<0$ implies
that there is a root of $D(\omega)$ on the upper half plane
(see Fig.\ref{fig:Nyquist}).

\end{document}